\documentclass[aps,prl,showpacs,twocolumn,floats]{revtex4}
\usepackage{amssymb}

\usepackage{amsmath}
\usepackage{graphicx,psfig}
\usepackage{dcolumn}
\usepackage{bm}

\input{tcilatex}
\begin{document}

\title{Thermal fluctuations and longitudinal relaxation of single-domain
magnetic particles at elevated temperatures}
\author{D. A. Garanin$^1$ and O.\ Chubykalo-Fesenko$^2$}
\affiliation{Institut f\"{u}r Physik, Johannes
Gutenberg-Universit\"{a}t, D-55099 Mainz, Germany}
\affiliation{Instituto de Ciencia de Materiales de Madrid, CSIC,
Cantoblanco, 28049, Madrid, Spain}
\date{\today}
\begin{abstract}
We present numerical and analytical results for the swiching times of
magnetic nanoparticles with uniaxial anisotropy at elevated temperatures,
including the vicinity of $T_{c}.$ The consideration is based in the
Landau-Lifshitz-Bloch equation that includes the relaxation of the
magnetization magnitude $M$. The resulting switching times are shorter than
those following from the naive Landau-Lifshitz equation due to (i)
additional barrier lowering because of the reduction of $M$ at the barrier
and (ii) critical divergence of the damping parameters.
\end{abstract}

\maketitle

%\pacs{75.60 Nt, 75.40. Mg, 76.20. +q, 75.50 Ss}

%\bibliographystyle{revtex}

%%%%%%%%%%%%%%%%%%%%%%%%%%%%%%%%%%%%%%%%%%%%%%%%%%%%%%%%%%%%%%%%%%%%%%%
The theory of thermal fluctuations of small magnetic particles is one of the
fundamental issues of modern micromagnetics. The conditions at which the
particle becomes superparamagnetic define the thermal stability of the
magnetized state and, therefore, is also valuable for technological
application such as magnetic recording \cite{Weller}. The basis of the
theory has been introduced by Brown \cite{Brown} who suggested to include
thermal fluctuations into the Landau-Lifshitz (LL) dynamical equation as
formal random fields whose properties are defined by the equilibrium
solution of the correspondent Fokker-Planck (FP) equation. He also derived
the Arrhenius-N\'{e}el formula to describe the relaxation rate in the
axially symmetric case of single-domain particles which was lately
generalized to the presence of external field \cite{Aharoni-Coffey,Kennedy}.
Since the paper of Chantrell and Lyberatos \cite{Lyberatos1}, this
Langevin-dynamics approach of Brown has been brought to numerical
micromagnetics to model the thermal properties of an ensemble of interacting
particles and, more generally, of ferromagnetic materials, interpreting the
micromagnetic discretization elements as small particles. Generally
speaking, the LL equation used in these simulations is a low-temperature
approximation only. Recently a generalized Landau-Lifshitz-Bloch (LLB)
equation for a ferromagnet \cite{Panina,LLB} was derived which is valid for
all temperatures and includes the longitudinal relaxation. The deviations of
the LLB dynamics from the LL dynamics should be pronounced at high
temperatures, especially close to the Curie temperature $T_{c}$. The
validity of this approach has been confirmed by the measurements of the
domain-wall mobility in crystals of Ba-and Sr-hexaferrites close to $T_{c}$
\cite{LLB-DW}.

Since the proposal of the heat-assisted magnetic recording (HAMR) \cite{HAMR}
the problem of high-temperature thermal magnetization dynamics has become of
large practical importance. The basic idea of HAMR is to write bits of
information at elevated temperature (close to the Curie temperature, where
the switching field is small) and store the information at room temperature.
To achieve a significant areal density advantage, the use of high-anisotropy
intermetallics such as $Ll_{0}$ FePt has been suggested \cite{Sun}.
Therefore, from both fundamental and applied points of view it is necessary
to consider the micromagnetics of small particles (or magnetic grains) at
elevated temperatures. The straightforward approach \cite{Lyberatos2} uses
the formalism of the standard stochastic LL equation, however with the
temperature-dependent parameters, i.e., the equilibrium magnetization $%
M_{e}(T)$ introduced through the mean-field approximation (MFA) involving
the Brillouin function, and the uniaxial anisotropy $K(T)$ through the
scaling relation $K(T)\sim M_{e}^{2}(T)$. However, this approach becomes
invalid at elevated temperatures as it does not incorporate the longitudinal
relaxational effects. The purpose of this Letter is to introduce the
theoretical formalism of the thermal fluctuations of single-domain particles
on the basis of the LLB equation which should be valid at all temperatures.
As a practical example, we consider analytically and numerically the thermal
switching of a FePt particle and discuss the conditions at which the
differences between the two formalisms emerge.

We start with the LLB equation \cite{LLB,LLBS} augmented by the white-noise
fields $\mathbf{\zeta ,}$ $\mathbf{\zeta }_{1},$ and $\mathbf{\zeta }_{2}$
in the form
\begin{eqnarray}
&&\mathbf{\dot{n}}=\gamma \lbrack \mathbf{n}\times \left( \mathbf{H}_{%
\mathrm{eff}}+\mathbf{\zeta }\right) ]+\frac{\gamma \alpha _{1}}{n^{2}}(%
\mathbf{n\cdot }\left( \mathbf{H}_{\mathrm{eff}}+\mathbf{\zeta }_{1}\right) )%
\mathbf{n}  \notag \\
&&\qquad {}-\frac{\gamma \alpha _{2}}{n^{2}}[\mathbf{n}\times \lbrack
\mathbf{n}\times \left( \mathbf{H}_{\mathrm{eff}}+\mathbf{\zeta }_{2}\right) %
]],  \label{LLB}
\end{eqnarray}
where $\mathbf{n\equiv M/}M_{e}(T),$ is the reduced magnetization, $\gamma $
is the gyromagnetic ratio, $\alpha _{1}$ and $\alpha _{2}$ are dimensionless
longitudinal and transverse damping parameters. The effective field $\mathbf{%
H}_{\mathrm{eff}}$ is given by
\begin{equation}
\mathbf{H}_{\mathrm{eff}}=-\frac{\delta F}{\delta \mathbf{M}}=\mathbf{H}+%
\mathbf{H}_{A}+(M_{e}/\chi _{\Vert })\left( 1-n^{2}\right) \mathbf{n,}
\label{HeffDef}
\end{equation}
where $F$ is the free energy density of the single-domain particle (cf. Ref.
\cite{LLB}), $\mathbf{H}$ and $\mathbf{H}_{A}$ are applied and anisotropy
fields, and $\chi _{\Vert }=\partial M_{e}/\partial H$ is the longitudinal
susceptibility. Parameters $M_{e},$ $\chi _{\Vert },$ and $\alpha _{1,2}$
depend on temperature and they are singular at $T_{c}.$ Within the MFA-based
model, one can use Eq. (4.9) of Ref. \cite{LLBS} with $K_{1}=K_{2}$ and $%
\gamma _{1,2}\Rightarrow \alpha _{1,2},$ rearranged to the form similar to
that of of Ref. \cite{LLB}:
\begin{equation}
\alpha _{1}=\frac{\lambda }{m_{e}}\frac{2T}{3T_{c}}\frac{2q}{\sinh (2q)}%
,\quad \alpha _{2}=\frac{\lambda }{m_{e}}\left[ \frac{\tanh q}{q}-\frac{T}{%
3T_{c}}\right] .  \label{alphas}
\end{equation}
Here $\lambda $ is a microscopic damping parameter that is temperature
dependent but noncritical at $T_{c}$,
\begin{equation}
m_{e}\equiv M_{e}(T)/M_{e}(T=0)  \label{meDef}
\end{equation}
is the reduced magnetization, and $q=3T_{c}m_{e}/[2(S+1)T]$. One can see
that in the vicinity of $T_{c}$ the relaxational parameters diverge: $\alpha
_{1}\cong \alpha _{2}\varpropto 1/M_{e}(T)$. In accordance with this
theoretical prediction, ferromagnetic resonance measurements on permalloy
have shown sharp increase of the damping close to the Curie temperature \cite
{Puzlei}.

The stochastic fields in the LLB-Langevin equation above can be, in fact,
introduced in many different ways. For instance, one can consider all three
fields as uncorrelated, set some of them to zero, or identify some of them
with each other. In all these cases one obtains different LLB-Langevin
equations and different stochastic trajectories. The physical quantities
obtained by averaging over realizations of $\mathbf{\zeta ,}$ $\mathbf{\zeta
}_{1},$ $\mathbf{\zeta }_{2}$ are, however, the same for all models. The
reason is that in all cases one obtains the same Fokker-Planck equation
(FPE), as was shown in Ref. \cite{Garcia-Palacios} for the LL-Langevin
equation. The FPE corresponding to Eq. (\ref{LLB}) can be obtained in the
same way as that for the LL equation (see Appendix of Ref. \cite{LLB}). The
result has the form of the conservation law
\begin{equation}
\frac{\partial f}{\partial t}+\frac{\partial }{\partial \mathbf{n}}\cdot
\mathbf{J}=0,  \label{FPE-LLB-Red}
\end{equation}
where $f\equiv f(\mathbf{n,}t)$ is the probability density and the
probability current $\mathbf{J}$ reads
\begin{eqnarray}
\mathbf{J} &=&[\mathbf{n}\times \mathbf{H}_{\mathrm{eff}}]f+\frac{\alpha _{1}%
}{n^{2}}\mathbf{n}\left( \mathbf{n\cdot }\left( \mathbf{H}_{\mathrm{eff}}-%
\frac{T}{VM_{e}}\frac{\partial }{\partial \mathbf{n}}\right) \right) f
\notag \\
&&-\frac{\alpha _{2}}{n^{2}}\left[ \mathbf{n}\times \left[ \mathbf{n}\times
\left( \mathbf{H}_{\mathrm{eff}}-\frac{T}{VM_{e}}\frac{\partial }{\partial
\mathbf{n}}\right) \right] \right] f.  \label{JFPELLB}
\end{eqnarray}

We will illustrate the stochastic dynamics of single-domain magnetic
particles for the model with the $z$-uniaxial anisotropy,
\begin{equation}
F_{A}=\left( M_{x}^{2}+M_{y}^{2}\right) /(2\chi _{\bot }),  \label{FADef}
\end{equation}
where $\chi _{\bot }$ is the transverse susceptibility that is a constant
within the MFA. We use Eq. (\ref{FADef}) rather than $F_{A}=-M_{z}^{2}/(2%
\chi _{\bot })$ to make $T_{c}$ independent of the anisotropy and thus to
simplify our formalism. Eq. (\ref{FADef}) could be rewritten using a
generalization of the micromagnetic anisotropy $K$ as $F_{A}=\left( \nu
_{x}^{2}+\nu _{y}^{2}\right) K$ (or as $F_{A}=-\nu _{z}^{2}K$), where $%
\mathbf{\nu }$\ is the magnetization direction vector, $\mathbf{\nu }\equiv
\mathbf{M}/M$ (see Ref. \cite{LLB}). This is not helpful, however, within
the approach based on the LLB equation. The problem is that $K=M^{2}/(2\chi
_{\bot })$ is not a constant and even not a function of temperature (cf. $%
K=M_{e}^{2}(T)/(2\chi _{\bot })$ used in Ref. \cite{Lyberatos2}), since the
magnetization magnitude $M$ can change dynamically during the thermal escape
process. It is convenient to scale the free energy density as $F=\left(
M_{e}^{2}/\chi _{\bot }\right) \tilde{F}$ with $\tilde{F}$ given by\cite
{Kachkachi}
\begin{equation}
\tilde{F}=-\mathbf{n\cdot h+}\frac{1}{2}\left( n_{x}^{2}+n_{y}^{2}\right) +%
\frac{1}{4a}\left( 1-n^{2}\right) ^{2},  \label{Ftilde}
\end{equation}
where $\mathbf{h\equiv }\left( \chi _{\bot }/M_{e}\right) \mathbf{H}$ and $%
a\equiv 2\chi _{\Vert }/\chi _{\bot }.$ We also define the temperature
variable $\sigma $ similarly to Ref. \cite{Brown}:
\begin{equation}
VF/T\equiv 2\sigma \tilde{F},\qquad \sigma \equiv VM_{e}^{2}/\left( 2\chi
_{\bot }T\right) .  \label{betatilDef}
\end{equation}
We will restrict our consideration to the case $H=0.$ In this case the
minima of $\tilde{F}$ are located at $n_{x}=n_{y}=0,$ $n_{z}=\pm 1,$ and $%
\tilde{F}_{\min }=0.$ The saddle point of $\tilde{F}$ is $n_{z}=0$ and $%
n_{\bot }\equiv \sqrt{n_{x}^{2}+n_{y}^{2}}=n_{s},$ where
\begin{equation}
n_{s}=\left\{
\begin{array}{cc}
\sqrt{1-a}, & a\leq 1 \\
0, & a\geq 1,
\end{array}
\right. \quad \tilde{F}_{\mathrm{sad}}=\left\{
\begin{array}{cc}
\left( 2-a\right) /4, & a\leq 1 \\
1/(4a), & a\geq 1.
\end{array}
\right.   \label{nsRes}
\end{equation}
In the limit $T\rightarrow 0$ (i.e., $\chi _{\Vert }\rightarrow 0$ and thus $%
a\rightarrow 0)$ Eq. (\ref{Ftilde}) confines the vector $\mathbf{n}$ to the
sphere $n\equiv |\mathbf{n}|=1,$ and the standard formalism based on the LL
equation is recovered. At nonzero temperatures, $a>0,$ the magnetization
changes its magnitude. In our model this effect is maximal at the saddle
point where the magnetization vector is perpendicular to the easy axis. One
can visualize the trajectory of this vector in the process of thermal
activation, after averaging over fluctuations, as an ellipsis going via the
saddle point. At $T=T^{\ast }$ defined by $a=1$ the ellipsis degenerates
into a line. In the range $T^{\ast }\leq T<T_{c}$ one has $a>1,$ and the
situation is qualitatively different. Here $\mathbf{n}$ contracts preserving
its direction along the $z$ axis and goes through zero at the saddle point,
then it grows in the opposite direction. These scenarios are very similar to
the transformation of the domain wall structure with temperature \cite
{LLB-DW}. Obviously the process of thermal activation of single-domain
magnetic particles cannot be described on the basis of the LL equation at
elevated temperatures. The crucial process of the longitudinal relaxation is
captured by Eqs. (\ref{LLB})--(\ref{JFPELLB}) based on the LLB equation.

The escape rate in the case $T\ll \Delta U$ has the form
\begin{equation}
\Gamma =\Gamma _{0}\exp \left( -\frac{\Delta U}{T}\right) ,\qquad \frac{%
\Delta U}{T}=\frac{VF}{T}\equiv 2\sigma \tilde{F}_{\mathrm{sad}}.
\label{GammaGeneral}
\end{equation}
In addition to the dependence $\sigma \varpropto M_{e}^{2}(T)$ in Eq. (\ref
{betatilDef}) that is responsible for the barrier lowering at elevated
temperatures, there is another source of the barrier lowering described by $%
\tilde{F}_{\mathrm{sad}}$ in Eq. (\ref{nsRes}). In particular, the value of $%
\tilde{F}_{\mathrm{sad}}$ at $a=1$ is two times smaller than its
low-temperature value, $a\rightarrow 0.$ The prefactor $\Gamma _{0}$ in Eq. (%
\ref{GammaGeneral}) can be obtained by solving the FPE, Eq. (\ref
{FPE-LLB-Red}), similarly to the derivation in Ref. \cite{Kennedy}. For $%
a\lesssim 1$ the result depends on $\alpha _{2}$ only$,$ since in this case
the barrier is being crossed by the pure rotation of the magnetization
vector. For $a\gtrsim 1$ both the longitudinal and transverse relaxation
becomes important, and it is difficult to find an analytical solution to the
FPE. Fortunately, in this temperature range $\alpha _{1}$ and $\alpha _{2}$
given by Eq. (\ref{alphas}) already become very close to each other, so that
one can set $\alpha _{1}\Rightarrow \alpha _{2}$ everywhere. Then the
calculation yields
\begin{eqnarray}
\Gamma _{0} &=&\alpha _{2}\omega _{1}\sqrt{\frac{\sigma }{\pi }}\sqrt{\frac{%
1-n_{s}^{2}}{a}}\exp \left[ \frac{a\sigma }{2}\left( 1-\frac{1}{a}\right)
^{2}\theta (a-1)\right]   \notag \\
&&\qquad \times \text{erfc}\left[ \sqrt{\frac{a\sigma }{2}}\left( 1-\frac{1}{%
a}\right) \right] .  \label{Gamma0General}
\end{eqnarray}
Here $\theta (x)$ is the step function and $\omega _{1}=\gamma M_{e}/(2\chi
_{\bot })$ is the ferromagnetic-resonance frequency. For $a\lesssim 1$ the
total rate simplifies to
\begin{equation}
\Gamma \cong 2\alpha _{2}\omega _{1}\sqrt{\frac{\sigma }{\pi }}\exp \left[
-\sigma \left( 1-\frac{a}{2}\right) \right] .  \label{GammaBelow}
\end{equation}
This reduces to the Brown's formula $\Gamma =2\alpha _{2}\omega _{1}\sqrt{%
\sigma /\pi }e^{-\sigma }$ \cite{Brown} in the limit $a\rightarrow 0.$
Exactly at $a=1,$ Eqs. (\ref{GammaGeneral}) and (\ref{Gamma0General}) yield $%
\Gamma =\alpha _{2}\omega _{1}\sqrt{\sigma /\pi }e^{-\sigma /2}.$ Just below
$T_{c}$ according to Eqs. (\ref{Ftilde}) and (\ref{betatilDef}) one has $%
a\sigma \varpropto \left( T_{c}-T\right) ^{2\beta -\gamma },$ where $\beta $
and $\gamma $ are the critical indices for the magnetization and
susceptibility. Within the MFA $a\sigma $ remains finite at $T_{c},$ whereas
for more realistic models it diverges. It makes, however, little sense to
work out the appropriate limiting expressions for $\Gamma $ because near $%
T_{c}$ the high-barrier approximation $\Delta U\gg T$ becomes invalid. In
fact, the prefactor $\Gamma _{0}$ in Eq. (\ref{GammaGeneral}) does not
strongly depend on $a.$ The main difference of our result from the Brown's
formula with a temperature-dependent barrier, $\Delta U\varpropto
M_{e}^{2}(T)$ is described by the two factors: (i) additional lowering of
the barrier because of the non-rigid magnetization, Eqs. (\ref{nsRes}) and (%
\ref{GammaGeneral}); (ii) crytical divergence of the damping at $T_{c},$ Eq.
(\ref{alphas}).

Brown has obtained the $1/\sigma $ correction to the escape rate for $\sigma
\gg 1$ in the form $\Gamma =2\alpha _{2}\omega _{1}\sqrt{\sigma /\pi }%
e^{-\sigma }\left( 1-1/\sigma \right) $ \cite{Brown-2}. Within the LLB
approach finding this correction in the whole range $0\leq a\leq \infty $ is
a complicated task. For $a\lesssim 1$ the correction factor in Eq. (\ref
{Gamma0General}) simplifies to
\begin{equation}
\left[ 1-\frac{1}{2\sigma }\left( 1+\frac{1}{n_{s}^{2}}\right) \right] .
\label{CorrFactLLB}
\end{equation}

To illustrate the practical implication of our approach, we consider thermal
switching of a FePt particle (magnetic grain) at high temperature. The
LLB-Langevin equation, Eq. (\ref{LLB}) has been integrated numerically with $%
\mathbf{\zeta }=0,$ $\left\langle \zeta _{i}^{\nu }\right\rangle =0,$ and
\begin{equation}
\left\langle \zeta _{i}^{\mu }(t)\zeta _{j}^{\nu }(t^{\prime })\right\rangle
=\frac{2k_{B}T}{\gamma M_{e}\alpha _{i}}\delta _{ij}\delta _{\mu \nu }\delta
(t-t^{\prime }),  \label{fluctfield}
\end{equation}
where $i,j=1,2$ and $\mu ,\nu =x,y,z$.

The microscopic relaxation parameter $\lambda $ in Eq. (\ref{alphas}) has
been found analytically for a spin-phonon interaction\cite{LLBS}. However it
is difficult to obtain reliable theoretical results for $\lambda $ in
general, as well as to extract $\lambda $ from experiments. For our
illustration below we just set $\lambda =0.1$, neglecting its temperature
dependence. The values of $m_{e}(T)$ of Eq. (\ref{meDef}) can be measured or
obtained from the Curie-Weiss equation $m_{e}=B_{S}(m_{e}\tilde{\beta})$,
where $\tilde{\beta}\equiv S^{2}J_{0}/(k_{B}T)$, $B_{S}(x)$ is the Brillouin
function, and $J_{0}$ is related to the experimentally measured $T_{c}$ via $%
T_{c}=(1/3)S(S+1)J_{0}$ within the MFA. For FePt $T_{c}=750$ K, and the best
fit for $m_{e}(T)$\ is obtained with $S=3/2$ \cite{Lyberatos2}. For FePt we
take $M_{e}(T=0)=1100$ emu/cm$^{3}$, $K(T=0)=1.24\times 10^{7}$ erg/cm$^{3}$%
, so that $\chi _{\bot }=M_{e}^{2}(T=0)/\left[ 2K(T=0)\right] =0.0488$ Oe~cm$%
^{3}$/emu. In the same way, the longitudinal susceptibility $\chi
_{||}=\partial M/\partial H$ can be measured or found analytically from the
MFA:
\begin{equation}
\chi _{||}=\frac{v_{0}M_{e}^{2}(T=0)}{S^{2}J_{0}}\frac{\tilde{\beta}%
B_{S}^{\prime }(m_{e}\tilde{\beta})}{1-\tilde{\beta}B_{S}^{\prime }(m_{e}%
\tilde{\beta})},  \label{suscept}
\end{equation}
where $v_{0}=6.4\times 10^{-23}$ cm$^{3}$ is the unit-cell volume and $%
B_{S}^{\prime }(x)\equiv dB_{S}(x)/dx.$

To integrate the LLB-Langevin equation, the Heun numerical scheme \cite
{Garcia-Palacios} has been used. The physically reasonable interpretation of
the stochastical process is that in the sense of Stratonovich, as was first
stressed for the LL equation in Ref. \cite{Garcia-Palacios}. Lately, it has
been shown \cite{Berkov} that even a naive Euler scheme which yields the Ito
solution, would converge to the proper averaged physical value, if the
magnetization is normalized at each time step, reflecting the conservation
of the magnetization length. However, in the case of LLB equation the
magnetization length is also a stochastic fluctuating variable, so that the
choice of the integration scheme should explicitly include the Stratonovich
interpretation.

The spins were prepared in the state $n_{z}=-1,$ and mean first-passage time
(MFPT) time was evaluated as the time elapsed until the magnetization
reached the boundary value beyond the barrier, $n_{z}=0.5$. The exact
position of this boundary only slightly changes the MFPT. Alternatively, one
can set the boundary at the top of the barrier, $n_{z}=0$. In this case one
has to multiply the time by 2, since in 50\% of all realizations the spin
crosses the barrier and in 50\% of all realizations it falls back \cite
{Haenggi}. In the high-barrier case, $T\ll \Delta U,$ the MFPT should
coincide with the relaxation time $\Gamma ^{-1}$ following from the FPE.

For a comparison, we also solved the (naive) LL-Langevin equation with a
constant but thermally reduced magnetization length,
\begin{equation}
\mathbf{\dot{n}}=\gamma \lbrack \mathbf{n}\times \left( \mathbf{H}_{\mathrm{%
eff}}+\mathbf{\zeta }\right) ]-\gamma \alpha _{2}[\mathbf{n}\times \lbrack
\mathbf{n}\times \left( \mathbf{H}_{\mathrm{eff}}+\mathbf{\zeta }_{2}\right)
]],  \label{LL-Langevin}
\end{equation}
where $\mathbf{H}_{\mathrm{eff}}$ is given by Eq. (\ref{HeffDef}) without
the last term. The temperature dependence enters this equation, as in the
LLB case, via the scaling of the anisotropy energy with $M_{e}^{2}(T)$ [Eq. (%
\ref{FADef}) and Eq. (\ref{Ftilde}) without the last term]. The non-rigorous
derivation of Eq. (\ref{LL-Langevin}) starts with the equation $\mathbf{\dot{%
s}}=\gamma \lbrack \mathbf{s}\times \left( \mathbf{H}_{\mathrm{eff}}+\mathbf{%
\zeta }\right) ]-\gamma \lambda \lbrack \mathbf{s}\times \lbrack \mathbf{s}%
\times \left( \mathbf{H}_{\mathrm{eff}}+\mathbf{\zeta }_{2}\right) ]]$ for
the spin vector of unit length $\mathbf{s}$. [The same starting equation is
used for the derivation of the LLB equation, Eq. (\ref{LLB}), in the
classical case]. Replacing $\mathbf{s}$ in this equation by its thermal
average, $\mathbf{s\Rightarrow m\equiv }\left\langle \mathbf{s}\right\rangle
,$ and rescaling $\mathbf{m}$ as $\mathbf{m\equiv n}m_{e}$ yields Eq. (\ref
{LL-Langevin}) with $\alpha _{2}=\lambda m_{e}.$ The latter is in a striking
contradiction with the rigorous LLB expressions for the damping parameters,
Eq. (\ref{alphas}). This difference becomes very important at elevated
temperatures, and there is no easy way to improve the naive LL results. In
our simulations of Eq. (\ref{LL-Langevin}) we just use the constant $\lambda
=0.1$ instead of $\alpha _{2},$ to conform with existing publications, e.g.,
with Ref. \cite{Lyberatos2}. Using $\alpha _{2}=\lambda m_{e}$ leads to even
more pronounced difference between the LL and LLB results.

\begin{figure}[t]
\label{Fig-Gamma} \centerline{\psfig{file=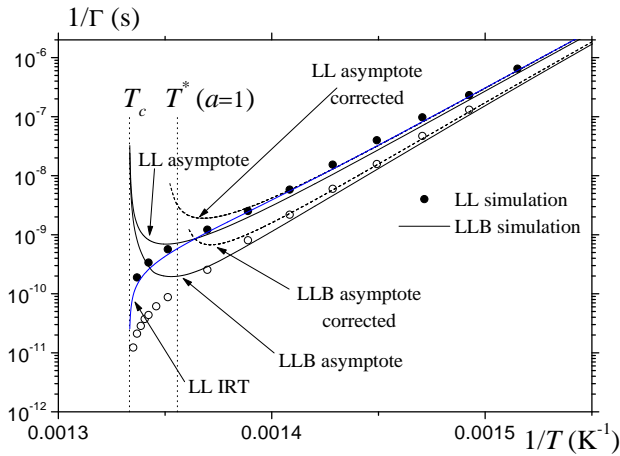,angle=-90,width=9cm}}
\caption{Switching times for a FePt particle with 8nm diameter calculated
numerically from the LL-Langevin and LLB-Langevin equations and analytically
from the approproate Fokker-Planck equations. While the integral relaxation
time (IRT) works very well at all temperatures, the high-barrier asymptotes
break down for $\protect\sigma \lesssim 1.$ The $1/\protect\sigma $
corrections improve the asymptotes for $\protect\sigma \gtrsim 1.$}
\label{Fig-Gamma}
\end{figure}

Fig. \ref{Fig-Gamma} shows the MFPT of a 8 nm one-domain FePt particle as a
function of temperature, calculated numerically from the LL-Langevin and
LLB-Langevin equations. These numerical results are compared in Fig. \ref
{Fig-Gamma} with Brown's analytical formula for the relaxation time $\Gamma
^{-1}$ \cite{Brown} and the result of Eq. (\ref{Gamma0General}). The
switching time calculated within the LLB approach is always lower than that
of the LL approach due to the additional lowering of the energy barrier (\ref
{nsRes}) and the critical growth of the damping at $T_{c}$. We have also
shown the temperature $T^{\ast }\simeq 738$ K at which $a=1$ and the
geometry of the barrier changes. For a given particle's volume, our
high-barrier approximation leading to Eq. (\ref{GammaGeneral}) becomes
invalid for $T\gtrsim T^{\ast }.$ We cannot increase the volume, however,
without violating the single-domain requirement.

Both Brown's formula for the LL model and Eq. (\ref{GammaGeneral}) for the
LLB model describing the Arrhenius dynamics are only valid for $T\ll \Delta
U $. Switching times showing an unphysical divergence near $T_{c}$ is the
signature of their breakdown. For the LL model, there is a better analytical
approach describing the thermally activated escape in terms of the integral
relaxation time (IRT) \cite{Panina,LLB,IRT} that is valid in the whole
temperature range. The IRT for the LL model is also plotted in Fig. \ref
{Fig-Gamma} showing a good agreement with the numerical data at all
temperatures. The possibility to find the IRT analytically results from the
spatial one-dimentionality of the FPE in the axially-symmetric case: $%
f=f(\theta ,t).$ For the LLB model there are two spatial coordinates, $%
\left\{ \theta ,n\right\} $ or $\left\{ n_{z},n_{\bot }=\sqrt{%
n_{x}^{2}+n_{y}^{2}}\right\} ,$ and a rigorous analytical solution for the
IRT seems to be impossible.

In conclusion we have introduced the formalism of the temperature
fluctuations within the mean field approach based on the
Landau-Lifshitz-Bloch equation. The new Langevin equation could serve as a
basis for the temperature-dependent micromagnetic approach for small
particles (or discretization elements) at high temperature, similar to
standard temperature-dependent micromagnetics but valid in the whole
temperature range. This new micromagnetics will have practical importance
for the heat-assisted magnetic recording applications.

%%%%%%%%%%%%%%%%%%%%%%%%%%%%%%%%%%%%%%%%%%%%%%%%%%%%%%%%%%%%%%%%%%%%%%

\end{document}